\documentstyle[12pt,epsf]{article}    % Specifies the document

\begin{document}

\baselineskip=25pt

\begin{center}{
{
\Large
Interacting Circular Nanomagnets}\\

\vskip 24pt
{\large A. A. Fraerman, S. A. Gusev, L. A. Mazo, I. M. Nefedov,
Yu. N. Nozdrin, I. R. Karetnikova, M. V. Sapozhnikov,
I. A. Shereshevskii, L. V. Sukhodoev}\\
\vskip 12pt
Russian Academy of Science\\
Institute for Physics of Microstructures\\
\vskip -3pt GSP-105, Nizhny Novgorod, 603600, Russia\\
E-mail: msap@ipm.sci-nnov.ru}
\end{center}
\vskip 2mm

\vskip 20mm
\begin{center}
Abstract
\end{center}

Regular 2D rectangular lattices of permalloy nanoparticles (40 nm in
diameter) were prepared by the method of the electron lithography. The
magnetization curves were studied by differential Hall magnetometry
for different external field orientations at 4.2K and 77K. The shape of
hysteresis curves indicates that there is magnetostatic interaction
between the particles. The main peculiarity of the system is the
existence of remanent magnetization perpendicular to easy plain. By
numerical simulation it is shown, that the feature of the magnetization
reversal is a result of the interplay of the interparticle interaction
and the magnetization distribution within the particles (vortex or
uniform).


\section{Introduction}

Now it is well known that competition between magnetostatic and exchange
energy in a very small ($\sim$20 nm) particle leads to a single - domain
state.  If the radius of a particle is sufficiently large, nonuniform
distribution of magnetization has minimal energy. It is a vortex in
isotropic magnetic disk. Such particle was referred to as circular
nanomagnet \cite{cowburn_99}. In the case of the individual nanoparticle
the vortex has perpendicularly magnetized core \cite{shinjo_00}.
Nowadays the vortex distribution of magnetization in nanomagnets is
under detailed investigation \cite{schneider_00}. Such systems are
considered as perspective for use in RAM (Random Access Memory)
device \cite{bussmann_99}.  It is obvious that distribution of
magnetization in nanomagnets depends on the interaction between particles.
The fundamental type of interaction, which can lead to the long-range
ordering \cite{cowburn_99_1} and to collective behavior in the system of
particles, is the magnetostatic interaction.  On the other hand, the
energy and a character of the magnetostatic interaction is
determined by the magnetization state of the particles.

In this work we experimentally investigate magnetization curves of
regular rectangular lattices of permalloy nanoparticles for different
external field orientations. We demonstrate that the magnetization
distribution within single particle depends on the magnetization process
and external field orientation to the lattice axis. It is a result
of the interplay of the interparticle interaction and the single particle
state. The particles can be both at single-domain and vortex state at
zero field.  The appearance of the magnetization vortices leads to the
appearance of the remanent perpendicular magnetization while magnetizing
perpendicular to the easy axis of the system.  The competition
between the single-domain and vortex states in the system of two
magnetostatically interacted magnetic nanodisks has been numerically
studied.

\section{Experiment}

Arrays of magnetic particles were fabricated by the electron-beam
lithography. The main feature of our method is the usage of
fullerine as a resist for electron lithography. Small sizes of $C_{60}$
molecules and the ability of fullerine to modify its physical and
chemical properties under exposure of electrons allow to use this
material for a high-resolution nanofabrication. The capabilities of
$C_{60}$ as a negative e-resist have been recently demonstrated by
fabrication of 20-30 nm Si pillars \cite{tada_96}. The main steps of the
procedure of manufacturing permalloy nanoparticles are thin films
deposition, exposing by e-beam, development and two-stage etching. We
used a double-layer mask containing the $C_{60}$ film as a sensitive
layer and Ti film as a transmitting layer. Permalloy and Ti films were
prepared by pulse laser evaporation on the substrate at room
temperature.  Fullerine films were deposited by sublimating of a $C_{60}$
powder at temperature of 350 C in a vertical reactor with hot walls and
cooling holder for the substrate. Transmission electron
microscopy, selected area diffraction and X-ray diffraction of the metal
and fullerine films were carried out to check the crystalline structure
and the thickness of layers.  Maximum sizes of the metal
crystallites did not exceed 5 nm. The $C_{60}$ films have an amorphous
structure. The thickness of the magnetic layers was varied from 25
to 45 nm. The thickness of the masking films was 20 nm for the $C_{60}$
layer and 30 nm for the Ti film.

The fullerine was patterned in the JEM-2000EX electron microscope with a
scanning electron microscopy (SEM) mode by 200 kV e-beam. we had
possibility to change the e-beam diameter from 10 nm and over.  Utilizing
of the high-energy electron beam decreases the amount of backscattering
electrons and the shape of patterns becomes defined better. The doses of
electron beam irradiation were 0.05-0.1 C/$cm^2$, because it assures the
reproducibility and uniformity of patterns sizes.  Electron beam
irradiation of  $C_{60}$ films reduces the solvability of fullerine in
organic solvents. The most likely reason of the changes of the
solvability is electron induced polymerization of $C_{60}$ molecules
accompanied by partially graphitization \cite{tada_96,mikushin_97}. The
exposed samples were developed in the toluene during 1 min and then
patterns were transferred to the Ti layer by the plasma etching with
$CF_2Cl_2$ atmosphere. The last step of the fabrication of the magnetic
particles was the $Ar^{+}$ ion milling of permalloy films using this
double-layered mask.  Basically, the resistance of the $C_{60}$ films to
the ion milling is sufficient to use this single mask with little bit
greater thickness, but the double etch steps are necessary to ensure a
uniformity and reproducibility of the sizes of the particles. By
carefully monitoring the elemental composition of the samples by means of
EDS (energy dispersion spectroscopy), qualitative microanalysis and
checking up the morphology of particles by SEM we can better detect the
final points of the plasma etching and ion milling processes. However,
usually we did some overmilling at the last step, to prevent
presence of any magnetic substance between the prepared particles. One of
the SEM images of the arrays of the ferromagnetic particles is presented
at Fig.1.  The shape of the ferromagnetic particles is a disc, which
height equals the thickness of the initial permalloy film.

\begin{figure}[th]
\centerline{
\epsfxsize=7cm
\epsffile{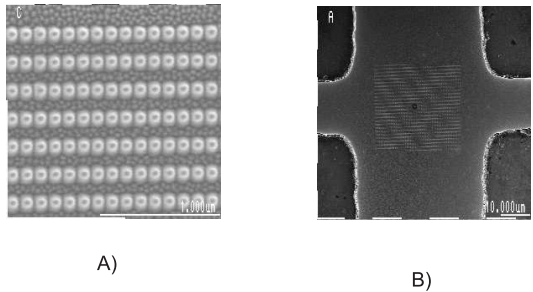}}
%\vskip -50mm
\caption[b]{
 The SEM-image of the sample 1 (See the Table). A) The lattice of the
40-50 nm particles is visible with the background of the 10 nm
roughnesses of the sublayer. B) The sample position in the Hall cross.  }
\end{figure}

The parameters of the investigated samples are summarized in the Table.
The symbols a and b denote the rectangular lattice parameters (the
distances between the centers of the particles), h is the height of the
particles and d is their diameter.  The total number of the particles is
equal to $10^5$.  \vskip 10mm

\begin{center}
Table\\
\begin{tabular}{c|c|c|c|c}
\hline
&&&&\\
N&a(nm)&b(nm)&h(nm)&d(nm)\\
&&&&\\
\hline
&&&&\\
1&90&180&45&50\\
&&&&\\
2&120&240&25&80\\
&&&&\\
\hline
\end{tabular}
\end{center}

\vskip 10mm

To carry out measurements of the magnetic properties we choose a
differential Hall microsensor technique. Recently it was shown that Hall
magnetometer is a very powerful tool for investigation of magnetic
properties of 2D nanoparticles lattices \cite{kent_94}. In our work we
applied a commercial magnetometer based on the Hall response in a
semiconductor (InSb) to investigate a collective behavior of the
fabricated 2D permalloy nanoparticle arrays. The widths of the current
and the voltage probes are 100$\mu$m and 50$\mu$m respectively  and the
thickness of the semiconductor layer is 10$\mu$m. The differential
magnetometer consisted of two Hall crosses with series connection of the
voltage probes. The lattice of the particles was produced in the active
area of one of the Hall crosses.  The difference in the Hall voltage
between this sample cross and the closely spaced empty cross is measured
using the bridge circuit \cite{kent_94}.  If the bridge is properly
balanced, the resulting output voltage is proportional to the sample
contribution to the perpendicular component of the magnetic field
induction. The large Hall response in the combination with the good
coupling of the small samples to the device results in the excellent spin
sensitivity (ratio signal/noise is approximately 100). The sensor works
over a large range of the magnetic field and temperature. We have
investigated the magnetic properties of the samples by measuring the
perpendicular magnetization as a function of the direction and magnitude
of the applied field. We have carried out the investigation for three
orientations of the external magnetic field:  1) the field is
perpendicular to the sample plane (The polar angle $\theta=0^{\circ}$);
2) the field is directed at $45^{\circ}$ to the sample plane along the
short side of the rectangle cell ($\theta=45^{\circ}$, the azimuthal angle
$\phi=0^{\circ}$); 3) the field is directed at $45^{\circ}$ to the sample
plane along the long side of the rectangle cell ($\theta=45^{\circ}$,
$\phi=90^{\circ}$). The direction $\phi=0^{\circ}$ corresponds to the
direction along the chains formed by the particles.  The chains are
elongated along the short side of the elementary rectangle (Fig. 1A).
The experimental results for the first sample for $T=4.2 K$ are
represented on the Figs.  2, 3, 4.

\begin{figure}[th]
\centerline{
\epsfxsize=7cm
\epsffile{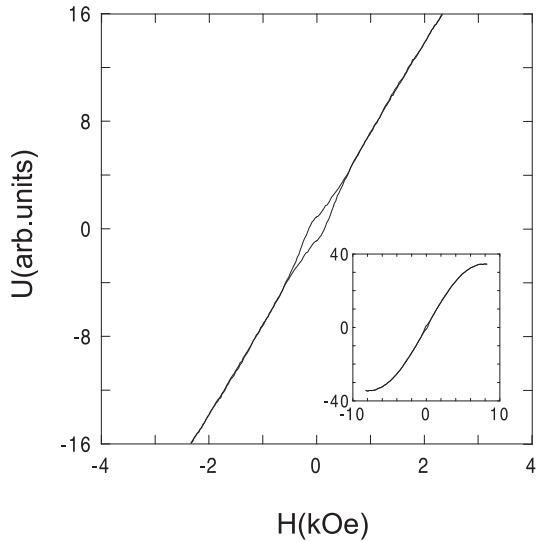}}
%\vskip -50mm
\caption[b]{The dependence of Hall
signal on the magnetic field with $\theta=0^{\circ}$.  The whole
magnetization curve is shown on the casing-in.  }
\end{figure}

\begin{figure}[th]
\centerline{
\epsfxsize=7cm
\epsffile{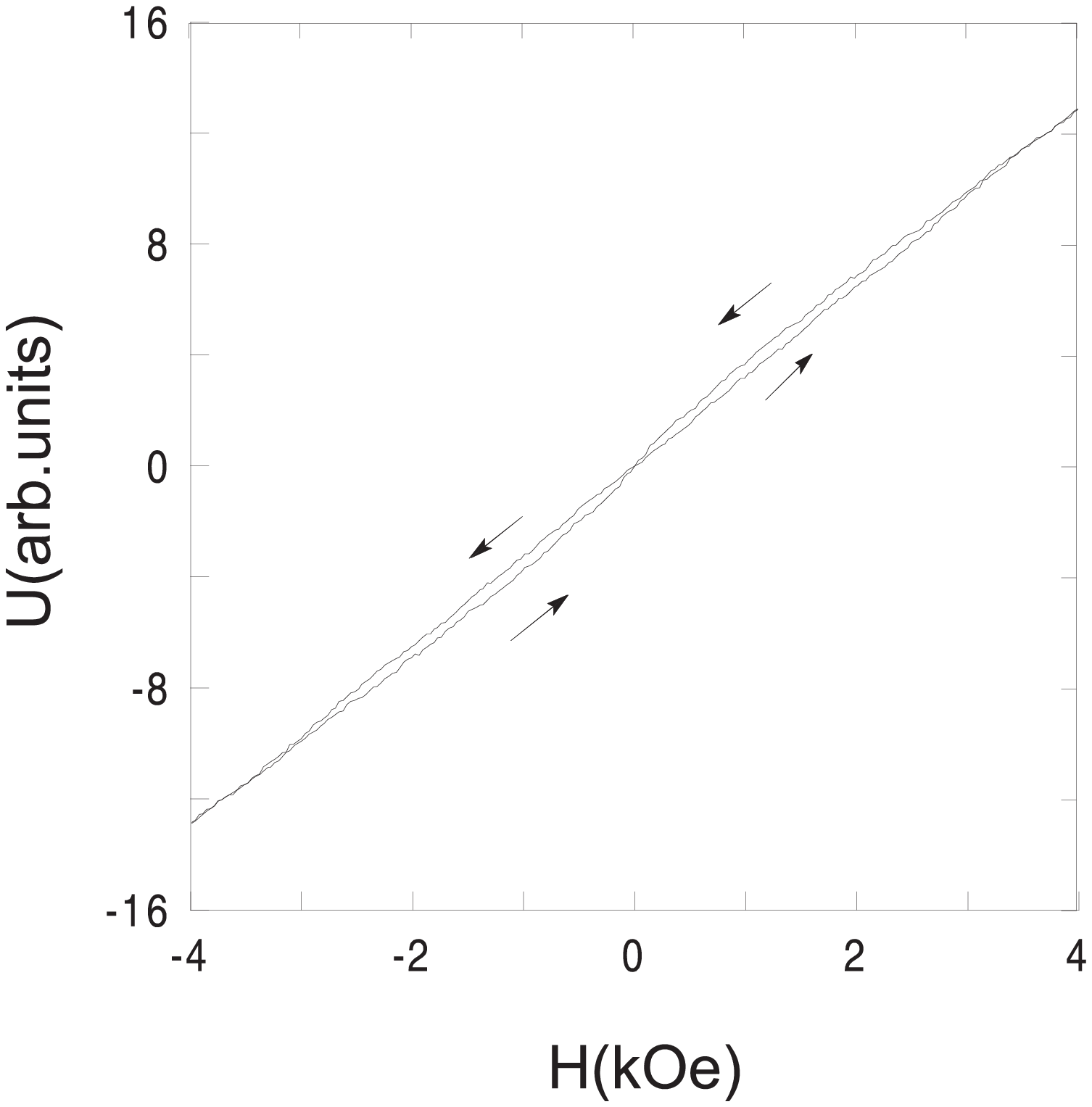}}
%\vskip -50mm
\caption[b]{
The dependence of the Hall signal  on the magnetic field with
$\theta=45^{\circ}$, $\phi=0^{\circ}$.
} \end{figure}

\begin{figure}[th]
\centerline{
\epsfxsize=7cm
\epsffile{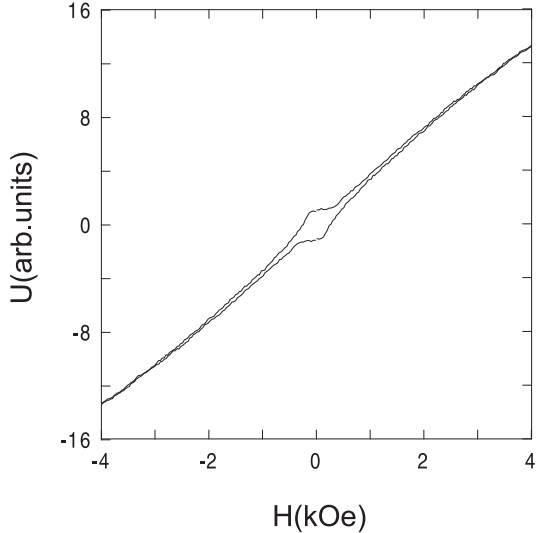}}
%\vskip -50mm
\caption[b]{
The dependence of the Hall signal  on the magnetic field with
$\theta=45^{\circ}$, $\phi=90^{\circ}$.
} \end{figure}

\noindent
The difference of the magnetization curves indicates the
collective behavior of the system. It is a result of the
magnetostatic interaction between the particles.  The hysteresis if the
field is directed at $\theta=45^{\circ}$, $\phi=0^{\circ}$ (Fig. 3) is
the attribute of the easy axis of the magnetization which is directed
along the short side of the rectangle cell. The perpendicular remanent
magnetization is absent in this case. The existence of such anisotropy in
the dipole system was discussed \cite{rozenbaum_91r_1,rozenbaum_91r}.
The magnetization curves if the field is directed at $\theta=0^{\circ}$
or $\theta=45^{\circ}$, $\phi=90^{\circ}$ (Figs. 2, 4) are qualitatively
similar. They have hysteresis in the weak magnetic field with the
remanent magnetization. Remanent magnetization is approximately 5 \% of
the saturation magnetization. The magnetization curves for the second
sample are qualitatively similar to those of the first sample, although
the samples have the different shape of the particles (See Table). The
particles have the polycrystal structure (this was determined by the
X-ray diffraction) and do not have the anisotropy of the form in the
plane of the system. In this case the difference of the magnetization
curves for the different orientation of the external magnetic field with
respect to the sample points to the collective behavior of the particles.
The existence of the anisotropy axis in the plane of the rectangular
lattice was discussed earlier \cite{rozenbaum_91r_1,rozenbaum_91r} and
was expected. As for the hysteresis and the remanent magnetization for
the sample with the rectangular lattice if the external magnetic field
direction is $\theta=0^{\circ}$ or $\theta=45^{\circ}$,
$\phi=90^{\circ}$, their existence seems to be unexpected.  The effect
can not be explained by the properties of single particle. Really, for
noninteracting particles the remanent magnetization does not depend on a
direction of the external magnetic field.

The first sample was also investigated at T=77K. The hysteresis if
the field was directed at $\theta=45^{\circ}$, $\phi=0^{\circ}$ did not
be observed. The hysteresis if the field was directed at
$\theta=0^{\circ}$ or $\theta=45^{\circ}$, $\phi=90^{\circ}$  was
qualitatively changed (Fig.  5).

\begin{figure}[th]
\centerline{
\epsfxsize=7cm
\epsffile{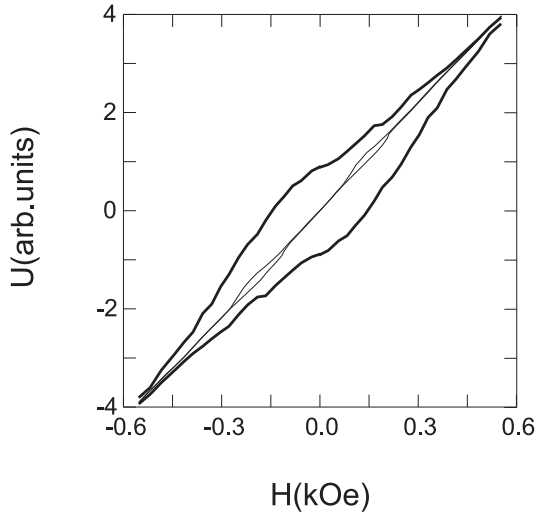}}
%\vskip -50mm
\caption[b]{
The changing of the magnetization curve hysteresis with the temperature:
the thick line for $4.2^{\circ}K$,  the thin one for $77^{\circ}K$. H is
directed at $\theta=0^{\circ}$.  } \end{figure}

We suppose that the observed behavior of the system is connected with
the fact that the particles of the examined sizes can be in two
states. First one is a single domain state, the second one is a vortex
distribution of the  magnetization. In the last case the core of the
vortex is magnetized perpendicularly \cite{shinjo_00}. The interplay of
the lattice anisotropy and the anisotropy of dipole interaction between
particles leads to the following behavior of the system. The particles
turn to be in the single domain state if the external field has the
component directed along the particle chains. In this case the
interaction within chain has the ferromagnetic character and therefore
stabilizes the single domain state.  On the other hand, if the system is
demagnetized within the field perpendicular to the chains the dipole
interaction has antiferromagnetic character within the chain and the
particles turn to be in the vortex state. As core magnetization has its
own coercivity, all cores are ferromagnetically ordered and the system
has the remanent perpendicular magnetization. We performed numerical
simulation to approve these suggestions.

\section{Numerical simulation}

Let us consider two magnetostatically interacting nanoparticles.
This model allows to investigate the magnetization
distribution within a particle in the anisotropic system with
interaction. We numerically solved the system of stochastic
Landau-Lifshitz equations. The effective magnetic field takes in
consideration the following components: the applied external magnetic
field, the demagnetization field of the system, the field of anisotropy,
is the exchange field within a particle and the random field defined by
the thermal fluctuations. The explicit Euler method was used to solve
such stochastic differential equation.  Note that such an approach to solve stochastic
LLH equation is well known and widely used in magnetic simulation
\cite{popko_99,boerner_97}. Besides we have used the
approximation of uniform distribution of the magnetization in the
perpendicular direction. Our numerical scheme is represented in detail in
our article \cite{ftt}.

Let us now discuss some results of numerical experiments. First of all,
for single magnetic disk we found that the vortex
magnetization state becomes ground state only when the height and radius
of disk exceeded some critical values.  This fact was pointed out
earlier in \cite{cowburn_98A} for the square-shape particles. Fig. 6
demonstrates an example of vortex state in cylindrical particle of
diameter 50nm and height 12.5nm at zero
external magnetic field. Note that such state has nonzero perpendicular
component of magnetic moment.  This can lead to appearance of hysteresis
loop in magnetization curve in systems of magnetic particles.

\begin{figure}[th]
\centerline{
\epsfxsize=7cm
\epsffile{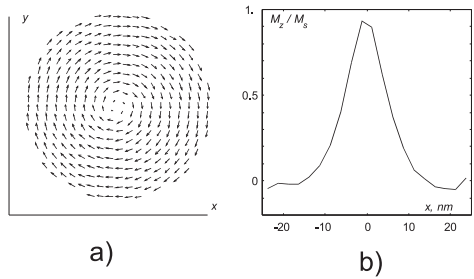}}
%\vskip -50mm
\caption[b]{
a) The vortex distribution of the magnetization in cylindrical particle.
b) the value perpendicular component of the magnetic moment.  }
\end{figure}

If there are two interacting particles, the magnetization curve is
changed.
Two different cases were considered:  first, when the projection of
the external magnetic field on disk plane lies along the pair of
the particles (it corresponds to $\theta=45^{\circ}$,
$\phi=0^{\circ}$ in experiment), and second, when this projection
lies perpendicularly to pair (it corresponds to $\theta=45^{\circ}$,
$\phi=90^{\circ}$ in experiment). Diameter of the particle
was 50nm, its height was 18nm, the distance between the particle centers
was 50nm.  The corresponding curves for the perpendicular magnetization
are shown in Fig.  7.  The hysteresis loop (stars in Fig.  7) is the
result of vortex penetrating into the particles at some value of the
applied field.  In contrast with this situation, the magnetization state
remains single-domain at all values of external field when its
projection lies along the pair of the particles, and as a result,
the curve for perpendicular magnetization has no a hysteresis loop
(Fig.  8).

\begin{figure}[th]
\centerline{
\epsfxsize=7cm
\epsffile{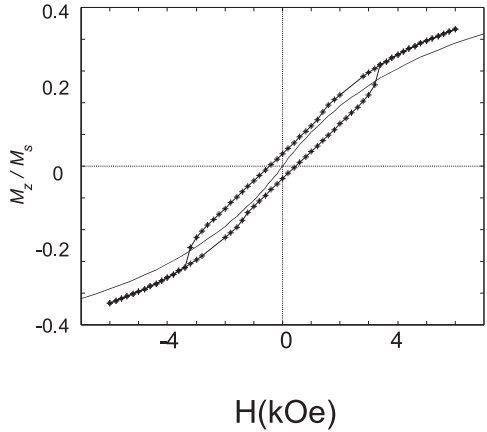}}
%\vskip -50mm
\caption[b]{
Perpendicular component of magnetization in two particles systems at
T=4K, when external field is applied along the pair of the particles
(solid line) and perpendicular to this direction (stars). Angle between
external field and the disc is $45^\circ$.  } \end{figure}

\begin{figure}[th]
\centerline{
\epsfxsize=8cm
\epsffile{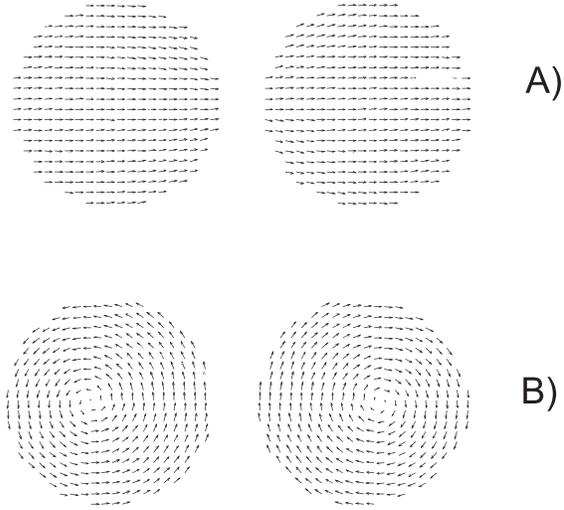}}
%\vskip -50mm
\caption[b]{
The uniform (A) and vortex (B) magnetization distribution in pair of the
permalloy discs.  } \end{figure}

\section{Discussion}

The following mechanism for the appearance of the remanent magnetization
implies from the results of experimental and numerical simulation
suggest with the external field orientation perpendicular to the particle
chain.  If the external magnetic field has component parallel to the
chains, the magnetostatic interaction between particles decreases the
total energy of the system and the remanent (H = 0) state is uniform and
the perpendicular component of magnetic moment is absent. In the case
when the external magnetic field is perpendicular to the chains,
magnetostatic interaction increases the total energy of the system. In
order to decrease the energy of the magnetostatic interaction,
distribution of the magnetization of the particle takes the vortex
form. As discussed earlier, a particle in the vortex state has a
nonzero magnitude of the remanent perpendicular magnetization.

Besides there were fundamental changes in the magnetization curve of
the sample observed when the sample temperature was
raised to 77K (Fig. 5). It points to the fact that thermal fluctuations
play a significant role in the system behavior, in spite of $T_c$ of bulk
permalloy is 885K.  It is possible that the absence of the remanent
magnetization at 77 K is caused by thermally - induced
switching between the vortex states with different polarization
\cite{gaididei_99}.  If the coercivity of the vortex core become less
than the antiferromagnetic magnetostatic interaction between cores the
remanent magnetization of the system will be equal to zero.  This
hypothesis requires further theoretical investigation.

So it is shown that the interaction in the regular 2D rectangular
lattice of nanoparticles plays a significant role. The interplay of
anisotropy of magnetostatic interaction and lattice anisotropy determines
the magnetization state of the particles. It can be both
vortex and single-domain state. Due to possibility of the vortex state
existence, the system can have the remanent magnetization directed
perpendicularly to the system plane.

$\qquad$ \section*{Acknowledgment.}
We are grateful to Prof. A.S.Arrott for helpful discussion. The work was
supported by the RFBR (00-02-16485) and PSSNS Program grants.

\bibliography{dipole}
\bibliographystyle{ms_phrev}

\end{document}